\documentstyle[12pt,aaspp,psfig]{article}
\slugcomment{\shortstack[r]{Accepted by MNRAS, 10th July 2000}}
\begin{document}
\newcommand\Msun {M_{\odot}\ }
\newcommand\Lsun {L_{\odot}\ }

\title{The Optical Velocity of the Antlia Dwarf Galaxy\footnotemark[1]}
\vskip0.5cm

\author{{\bf Eline Tolstoy\footnotemark[2] \footnotemark[3]}}
\affil{European Southern Observatory, Karl-Schwarzschild-str 2, 
D-85748, Garching bei M\"{u}nchen,
Germany}

\author{{\bf Michael Irwin\footnotemark[4]}}
\affil{Institute of Astronomy, University of Cambridge, 
Madingley Road, Cambridge, CB3 0HA, England, UK}

\footnotetext[1]{Based on observations collected at the European Southern
Observatory, Chile, in service mode, proposal number 63.N-0197}
\footnotetext[2]{Current address: 
UK GEMINI Support Group, University of Oxford, 
Nuclear and Astrophysics Laboratory, Keble Road, 
Oxford OX1 3RH, UK}
\footnotetext[3]{email: etolstoy@astro.ox.ac.uk}
\footnotetext[4]{email: mike@ast.cam.ac.uk}

\begin{abstract}

We present the results of a VLT observing program carried out in
service mode using FORS1 on ANTU in Long Slit mode to determine the
optical velocities of nearby low surface brightness galaxies. Outlying
Local Group galaxies are of paramount importance in placing
constraints the dynamics and thus on both the age and the total mass
of the Local Group.  Optical velocities are also necessary to
determine if the observations of HI gas in and around these systems
are the result of gas associated with these galaxies or a chance
superposition with high velocity HI clouds or the Magellanic
Stream. The data were of sufficient signal-to-noise to obtain a
reliable result in one of the galaxies we observed - Antlia - for
which we have found an optical helio-centric radial velocity of
351~$\pm$~15~km/s.

\end{abstract}

\keywords{GALAXIES: INDIVIDUAL: ANTLIA, GALAXIES: KINEMATICS AND DYNAMICS,  
GALAXIES: LOCAL GROUP}

\section{Introduction}

Several relatively isolated outlying members of the Local Group:
Tucana, Phoenix, Cetus and Antlia currently have no measured optical
radial velocities.  Although Phoenix (Carignan {\it et al.} 1991;
St-Germain {\it et al.} 1999), Antlia (Fouqu\'{e} {\it et al.} 1990)
and possibly Tucana (Oosterloo {\it et al.} 1996), have tentative HI
detections, there are as yet no compelling reasons to believe that any
of the detected HI is necessarily associated with the optical galaxy
({\it e.g.} Young \& Lo 1997).  These outlying members of the Local Group are
of paramount importance in determining membership and dynamics of the
Local Group and in particular for placing tight constraints on both
the age and total mass of the Local Group.  In the case of Phoenix an
optical radial velocity is needed to assess possible membership as the
outermost satellite system of our Galaxy; thereby constraining the
mass of the Galactic Halo out to the unprecedented distance of 400~kpc.

Optical radial velocities can be used to directly probe the
relationship between HI gas detections in and around dwarf spheroidal
galaxies (dSphs), coincident line-of-sight Galactic high velocity
clouds ({\it e.g.} Braun \& Burton 2000), or high velocity clouds associated
with the Magellanic Stream.  This is one of a number of open questions
in the attempt to establish if a link exists between dSphs with no gas
in their centres and gas rich, currently star-forming, dwarf irregular
galaxies.  Proving that gas is associated with dSphs would give
support to the idea that HI gas can be severely disrupted, perhaps
during a star formation episode, blown into an extended dark matter
halo, but still retained ({\it e.g.} Sculptor, Carignan {\it et al.} 1998).

However, obtaining accurate radial velocities of these faint and
diffuse objects has required the availability of Southern
8m-class telescopes.  In this paper we present the results of
VLT/FORS1 long slit observations of three of the isolated, outlying
members of the Local Group: Antlia, Cetus and Tucana (see Table~1).
All of these galaxies are close enough such that their stellar
populations are easily resolved, and none have previously measured
optical velocities.  Unfortunately, due to weather conditions and the
proximity of the Moon for two sets of observations, only the data for,
Antlia, was of sufficient signal-to-noise to give a reliable result,
but for completeness we have included an outline description of the
other data.

Antlia has been discovered more than once; in the radio, via an HI
survey by Fouqu\'{e} {\it et al.} (1990) and in the optical, where it
was first confirmed as a member of the Local Group by Whiting {\it et
al.} (1997).  Antlia is a companion galaxy (on the sky) to NGC~3109,
which is within the measurement errors at the same distance, and has a
comparable HI radial velocity (v$_{\odot}$(HI) = 361 km/s compared to
403 km/s for NGC~3109).  However, the HI gas has also been suggested
to be a chance superposition with a side-lobe of the HI detected in
NGC~3109 ({\it e.g.} Blitz \& Robishaw 2000).  More recent VLA data however
suggests that the HI is in fact associated with Antlia (Skillman,
private communication).  Measuring an optical velocity for Antlia
would not only clarify the whole situation but also add weight to the
hypothesis that most, if not all, dSphs have associated HI gas.

Tucana, which is also a relatively recently discovered member of the
Local Group (Lavery \& Mighell 1992), lies close to the line-of-sight
of the Magellanic Stream and as such it is hard to assess the
significance of the discovery of nearby HI gas by Oosterloo {\it et
al.} (1996).  Tucana is at a distance of 900~kpc, and lies on the
opposite side of the Local Group to the M31 sub-system.  Until the
discovery of Cetus, it was the only known unambiguously isolated dSph
in the Local Group, and therefore a unique probe of the galactic
evolution and internal dynamics of an isolated dSph.

Cetus is a newly discovered galaxy in the Local Group (Whiting {\it et
al.} 1999) about which very little is known.  Although located in the
general direction of the extension of the Local Group toward the
Sculptor group, Cetus appears to be an isolated dSph and as such may
join the class currently occupied only by Tucana. The first deep CMD
of Cetus, which reaches down to the level of the horizontal branch
(HB) shows this to be predominantly old system with a highly populated
HB and no signs of recent ($<$1Gyr old) star formation (Tolstoy {\it et
al.} 2000).  Preliminary attempts to detect Cetus in the HIPASS survey
(see Whiting {\it et al.} 1999) have revealed nothing, which suggests
that Cetus does not contain any HI gas. Therefore, the only way to
learn more about its position in the Local Group, and if it really is
an isolated system, is to measure an optical radial velocity.

\section{Observations}

The observations were obtained in service-mode, with UT1/FORS1 in the
Long Slit Spectroscopy (LSS) instrument set-up, between May and August
1999 (see Table~2).  The FORS1 long slit spans 6.8 arcmin and was used
with a slit width of 1 arcsec and the GRIS-600I+15 grism along
with the OG590 order-sorting filter, to cover the Ca~II triplet
wavelength region with as high resolution as possible.  With this
setting the pixel sampling is close to 1\AA \, per pixel and the
resolution $\approx$2-3\AA \, over the wavelength range 7050$-$9150\AA
\,. This is the maximum resolution that can be obtained with FORS1
without resorting to a narrower slit. Although this is a wavelength
range at which the FORS1 CCD (Tektronix) has reduced sensitivity it is
where the red giant stars we were aiming to detect are brightest. The
Ca~II triplet is also a useful unblended feature to accurately measure radial
velocities ({\it e.g.} Hargreaves {\it et al.} 1994) and there are
abundant narrow sky lines in this region for wavelength calibration
and/or spectrograph flexure monitoring.

Our observation request consisted of two different slit positions per
target galaxy, or in ESO parlance two separate Observation Blocks
(OBs), which could be (and indeed were) observed on different nights,
and in different conditions (see Table~2).  Although we did not have
very strict observing constraints (seeing $<$ 1arcsec, any weather),
most of our data were taken out of specifications, either because the
seeing was worse, or the moon constraint was violated. Two of the
three galaxies (Cetus and Tucana) in our sample were taken in full
moon (96\% and 100\% illumination) conditions and Cetus was only 30
degrees away from the moon. These poorer than anticipated conditions
with no corresponding increase in exposure time, combined with the
reduction in the sensitivity of the system from the initial values
obtained with the exposure time calculator, meant all the data ended up 
of much lower signal-to-noise than anticipated.

Each OB was made up of a target observation in conjunction with a
sequence of radial velocity (rv) standards (both individual K-giants
and bright globular clusters, see Table~3). The K-giant absolute rv
standards provided the basic velocity standardization in addition to
Ca~II triplet template spectra for comparison with the much lower
signal-to-noise dwarf galaxy spectra.  By observing bright globular
clusters with known radial velocities we further sought to test our
methodology with data more closely resembling the primary targets. A
useful by-product of this is additional template spectra to
cross-correlate with the main target objects.

Typically we observed at two different slit positions on each target
galaxy, and sometimes also on the calibration globular clusters.  We
attempted to align the slit to cover known, likely, stellar members
where possible, but for some systems we lacked suitable pre-imaging
exposures with a world coordinate system to use as input into the
FIMS, the tool used to make OBs for execution at the telescope.  In
the latter case it was a reasonable assumption that we would hit
several objects per slit as these galaxies are fairly densely
populated by (faint) stars.  We also planned to use the combined integrated
unresolved light of the underlying galaxy across the slit.

In the case of Antlia and Rup~106 one of our OBs was executed twice
(no explanation given), with the result that instead of two separate
target slit positions we had one. This had the (unanticipated, but
interesting) benefit that we could check the repeatability of our
results as these two OBs were observed one month apart.

\newpage
\section{Data Reduction and Calibration}

All basic calibration data provided for UT1+FORS1 observations, 
{\it i.e.} not only bias frames, but also flat-field frames and arcs for
wavelength calibration, are obtained during daylight hours.  Because
our observations require us to make very accurate determinations of
the wavelength solution and because we are typically looking at very
faint objects we were very careful in assessing the usefulness of the
(day-time) calibrations provided and the potential for spectrograph
flexure between arc calibration measurements and target measurements
on sky. We were seriously concerned about the possibility of
flexure of this Cassegrain-mounted spectrograph, between
the daytime wavelength calibrations and the nightly observations, and
also the long term effects since our data were taken over a period of
a few months, which could easily introduce systematic wavelength
offsets between data sets.

The default lamp flat fields provided included a multitude of spurious
artifacts at a variety of scale-lengths rendering them difficult to use
for sensitivity corrections. Similar spurious internal reflections
also made the arc calibrations frames difficult to use, since in
long-slit mode it is essential to accurately trace the spatial
curvature of the spectral lines and extract an accurate
two-dimensional wavelength solution.  We also had several basic
problems with the calibration data for the first OB (Antlia, taken in May
1999) because all the provided bias, flat-field and arc calibration
frames were taken using four-quadrant readout mode, whereas the
nightly observations used single-channel readout.

Because of these problems we made an executive decision to not use any of
the daytime calibrations to either flat-field or wavelength
calibrate the data.  Most of the difficulties with
the daytime calibration seem to stem from the reflections caused by
the calibration lamps which live within FORS1 and are shone out of the
instrument through the LADC (linear atmospheric dispersion corrector,
{\it i.e.} two large pieces of glass) 
and bounced off the (closed) beam shutter 
back through the LADC before making it to the spectrograph slit. Not
surprisingly there are reflections due to scattered light in the
resulting calibration data.

The problems with the provided calibration data forced us to adopt a
more creative approach than normal using the following steps all of
which were done using the IRAF package.
\footnote{IRAF is distributed by the National Optical Astronomy
Observatories, which are operated by the Association of Universities
for Research in Astronomy, Inc., under cooperative agreement with the
National Science Foundation.}

Fortunately the fixed pattern bias is quite stable and there were sufficient 
single readout bias frames for this CCD taken over this period in the ESO
archive.  These were used to construct a master bias frame which,
together with the under/overscan regions within the spectroscopic data,
was used to trim and bias-correct all the data.

In the ESO archive there were also I broad-band filter twilight flat-field 
frames available taken using the same CCD.  Since this covers the same 
wavelength region as the spectroscopic observations, we were able to use these
to produce a high quality flat-field frame which could be used on the
spectroscopic images.  Illumination gradients across the flat were
mapped using a combination of median and Gaussian filtering to produce
a smooth two-dimensional illumination map.  The stacked flat-field
frame was then divided by the smooth illumination map and the result
used to correct all the spectroscopic data for the pixel-to-pixel 
sensitivity variations/defects of the CCD in the standard manner.

Working in the wavelength range 7050--9150\AA \,, at a resolution of
1\AA /pixel, the night sky lines are plentiful and well distributed
over the wavelength range and could be used to directly map the
two-dimensional wavelength distortion and accurately calibrate the
spectra.  The adopted reference wavelengths were taken from the on-line
Keck LRIS skyline plots, which were in turn based on a compilation by 
Osterbrock \& Martel (1992).  The two-dimensional distortion corrections 
were found to be stable to $\approx$0.1\AA \, within an observing night, 
apart from small global translations (flexure) along the dispersion axis, 
and could therefore be used for the generic two-dimensional wavelength 
calibration of all the data for that night.  This could otherwise have
been a serious problem since the radial velocity standards do not have
any significant skylines visible due to their much shorter exposure times.
Additional checks and corrections for systematic wavelength shifts
were made using the final extracted spectra and are described later.

We stress again that since it was not allowed telescope practice to take 
calibration arcs before and after each integration to monitor the instrument 
flexure (advertised to be zero at any distance from the zenith), we 
used the sky lines in the final extracted spectra to both measure and correct 
for this flexure.  The variation in apparent sky line wavelengths indicated 
that although the flexure is small, and certainly never much bigger than a 
pixel, or $\sim$35~km/s, it is not negligible, and not surprisingly, it is 
strongly dependent on zenith distance (see Table~4).

Each target observation of a dwarf spheroidal galaxy consisted of
3$\times$1800s exposures.  There were no measurable internal shifts
found for frames within a sequence, so these were combined in a
straight-forward manner eliminating cosmic ray events, prior to
wavelength calibration, carried out using the night sky lines.
Globular cluster observations were typically integrations of 150$-$250s
which were treated in the same way as the target galaxy
observations.

The position of the slit across the central region of Antlia taken
from the FIMS file used to prepare the OBs is shown in Figure~1. Also
marked are four of the stars clearly detected in the Antlia spectra. The 
position of all the resolved objects detected in the Antlia long-slit spectra
are plotted on the Colour-Magnitude Diagram (CMD) of Antlia (from Tolstoy 
1999) in Figure~2.  From the location on the CMD, and also position on
the slit, stars B, C and D are all very likely to be members of Antlia, 
while A and E are much too bright to be associated with Antlia.  Object F 
appears to be a background galaxy in the slit.  The identification of all the 
resolved objects in the slit and their location on the CMD provides an 
excellent independent corroboration of the optical velocity determinations 
and was actually computed after the results from the spectroscopy were known.

\section{Extracting and Wavelength Calibrating the Spectra}

For the globular clusters and dSphs, image sections summed along
the spatial axis were used to identify resolved stellar, or background
galaxy images, in the slit and also to define the underlying
unresolved background of the dwarf spheroidal galaxy for spectroscopic
extraction.  Spectra for these objects plus the integrated background
were then extracted in the usual manner taking care to optimise the
region chosen to represent the local sky.  The computed sky spectrum
was also extracted contemporaneously for later use in checking the wavelength 
solution systematics (see Figure~3).  
For the radial velocity standards contributions to
systematics from slit centering of the extremely bright stars is also an issue
that limits the achieved final accuracy and is reflected in the scatter 
($\sigma_v \sim 5$km/s) of the derived velocities for the standards in Table~4.

In more detail:- The IRAF task IDENTIFY was used to make the wavelength 
calibration using the Keck LRIS sky-line identifications.  The reference
image used for this was always one of the coadded 3$\times$1800s dSph summed
frames on a nightly basis.  Next the task REIDENTIFY was used to trace the 
sky lines to track the two-dimensional distortion. FITCOORDS is then used to 
produce a two-dimensional polynomial distortion transformation
which was input to TRANSFORM, {\it i.e.} to straighten out, the long-slit 
distortions.  In all cases a single distortion mapping was stable enough to be 
used within a given nights observations. At this point the IRAF task APALL 
was used to extract both object spectra and sky spectra in the 
neighbourhood of each object in the usual manner.

The extracted one-dimensional sky spectra were then cross-correlated, using
task FXCOR, against the reference image sky spectra, after continuum removal 
and suitable apodizing, to measure the stability of the wavelength solution 
and hence determine the flexure of the spectrograph.  The cross-correlation 
of sky lines within a single two-dimensional frame gives negligible velocity
shifts of $\pm$3 km/s which is comparable to the juxtaposition of the
wavelength calibration errors and correlation-function estimation errors.

Monitoring the wavelength stability of the radial velocity standards was more
challenging.  The globular clusters Rup~106, NGC~6752 and Pal~12, provided a 
link between the deeper sky exposures of the target dSphs and the radial 
velocity standards where no sky emission lines are visible.  First the 
globular cluster extracted object sky spectra were cross-correlated with the 
dSph reference spectra to place them on the same zero-point system with an 
accuracy of $\approx$3~km/s.  Then using a Gaussian fit to the lower half of 
the atmospheric A-band sky absorption features in the globular cluster spectra
and the radial velocity standards spectra, the wavelength system of the radial
velocity standards could be tied to the globular cluster spectra.  The error 
in the location of the A-band was found to be $\approx$5~km/s using this 
fitting procedure, when there was good signal-to-noise.  After adopting this 
bootstrapping of the wavelength solution, a series of internal consistency 
checks indicated the overall final systematics of the wavelength solution 
are in the range 5-10~km/s, more than adequate for the current goal.

The largest flexures found amounted to about 1 pixel and were found between
the two Antlia data sets taken on different nights one month apart
and for one of the radial 
velocity standards taken at an airmass of 2.4 on the same night as the
primary wavelength calibration data (see Table~4).

\section{Determining the Radial Velocities}

In the previous section we described how we have extracted spectra and
carried out a number of tests to determine the errors due to the 
wavelength calibration and correct for them.  Now we can make an
accurate comparison of the velocities of the lines in the Ca~II triplet 
in all our observations and be confident that any offsets 
between the radial velocity standards, the cluster observations and dSphs
are all representations of the true velocity differences (limited by
signal-to-noise) between these objects and thus allow us to estimate
the true optical velocity of the target galaxies.

One of the radial velocity standards, HD~107328 of spectral type K1~III, was 
used as the primary cross-correlation template.  First, an order 12 cubic 
spline continuum fit was made and subtracted from the template spectrum.
Then the IRAF SPLOT package was used to
snip out featureless regions from the
continuum-subtracted template and to also remove the sky A-band absorption 
feature at $\approx$7600\AA \, by setting these regions to zero.  This gives
unit weight to those spectral regions
with maximal information 
for radial velocity determination ({\it i.e.} 
unambiguous stellar absorption
features) and zero weight to all other regions.  
Even the regions
between the Ca~II triplet lines were set to zero to minimise problems with 
residual sky line features causing zero velocity cross-correlation locking.

Radial velocities were then measured for all the spectra, using the standard 
Fourier cross-correlation package in IRAF, FXCOR.
An example of the results for Antlia is shown 
in Figure~4.  The measured radial velocity was then corrected for: the 
template radial velocity offset; the topo-centric correction to a helio-centric
system; and for flexure.  All of the measurements, the corrections and the 
final values for the individual objects extracted from the spectra are listed 
in Table~4, including 
Antlia, the radial velocity standards and the globular clusters 
Rup~106 and NGC~6752.  The results for Pal~12, Cetus and Tucana 
were of such poor quality that no useful information on their radial 
velocities could be obtained at all.

For all the globular cluster and dSph observations, individual resolved 
objects were extracted separately in an attempt to identify system members 
and possible foreground star, or even background galaxy, interlopers.  
In addition to the individual objects in Antlia we also extracted the majority
of the unresolved flux, plus central objects, in an attempt to boost the 
signal-to-noise.  Three of the extracted spectra for Antlia (b,c,d) had radial 
velocities that matched within the errors and were also found later to lie
at the tip of the red giant branch (see CMD in Figure~2).  The summed spectra
for Antlia stars B, C, D in the Ca~II triplet region is shown in Figure~3,
together with the vertically offset spectrum for NGC~6752 and an example
sky spectrum.  The Ca~II triplet features in Antlia are clearly visible, 
redshifted by about 10\AA \, from those of NGC~6752.  Note also how the strong
night sky lines, plotted scaled down by a factor of 75 with respect to the 
Antlia spectrum, cause residual sky features even in the summed Antlia 
spectrum, emphasizing the crucial importance of careful template construction 
to avoid sky line residual locking.  The final result for Antlia is an
optical helio-centric radial velocity of 351$\pm$15~km/s.  The error estimate
includes a contribution from the {\it rms} FXCOR estimation errors and the
systematics of the whole wavelength calibration procedure, which contribute
in roughly equal measure.

The same analysis was carried out identically for Tucana and Cetus.
In both these cases the extremely bright sky lines and poor signal-to-noise
due to full moon make the extracted spectra too poor to be of any use.

\newpage
\section{Conclusions}

It is clear from the results of the individual stars observed in
Antlia (and also in Rup~106 and NGC~6752) when we pick up a star which is
member of the system and when we find a foreground, or in the case of
Antlia-1-F a background galaxy.  The previously acquired VLT Antlia images
were used to select two candidate stars to lie on the slit (B,C), whilst 
(A, D, E) serendipitously turned up in the slit.
The three stars B, C, and D are consistent with membership of Antlia
both on account of their position in a Colour-Magnitude Diagram and
their consistent optical velocities which are not likely to be similar
to any objects of similar colour and magnitude in our galaxy.

In more detail, if the stars B, C and D in Antlia were Galactic and in
the disk they would have a radial velocity with respect to the LSR of
$\sim$360 km/s and if they are Halo objects the galactocentric radial
velocity is more appropriate at $\sim$150 km/s.  Contamination by disk
K-dwarfs seems highly unlikely on two counts: the extremely high
LSR-corrected radial velocity; and the distance any putative K-dwarfs
would have to be.  At V$\sim$23 a typical K-dwarf would have a distance
modulus of $\sim$15 and would have to be at around 10~kpc distance and hence
$\sim$4~kpc above the Galactic Plane.  In the case of possible Halo star 
K-giant contamination; the radial velocity is high but still around the 
2-$\sigma$ Halo velocity dispersion.  However, we can rule out K-giants due 
to distance constraints, since a typical K-giant at V$\sim$23 would have a 
distance modulus over 25 and hence be at a distance of order 1 Mpc or more.  
The only other possible contaminants are K-dwarfs from the Halo/Spheroid.  
However, at a galactocentric distance of $\sim$14~kpc, the space density of
Halo/Spheroid K-dwarfs at V$\sim$22 is $\approx$100 per square degree
({\it e.g.} Bahcall, Schmidt \& Soneira 1983).  With the caveat that
aposteriori statistical estimates should only be a guide at best, the central 
part of the slit covers some 60 arcsec$^2$, implying a probability of
finding three or more Halo/Spheroid K-dwarfs in this part of the slit by 
chance, of a few parts in ten thousand.  Factoring in the probability of the 
agreement in velocities, leads to a combined probability of less than one 
part in a million of this being a random contaminating Galactic sample.
Since Antlia is also 
located well away from any other sources of possible near-Galactic
contamination such as the Saggitarius Dwarf tidal stream, or the Magellanic 
Stream, we conclude that stars B, C and D are members of Antlia and that
hence the optical helio-centric velocity of Antlia is 351$\pm$15~km/s.

Our determination of the optical velocity for Antlia closely matches
the HI velocity for this galaxy (Fouqu\'{e} {\it et al.} and Skillman,
private communication).  Thus our results combined with those of Fouqu\'{e} 
{\it et al.} unequivocally show that Antlia contains modest amounts, 
($\approx 8\times10^5 \Msun$), of HI gas centered on the optical
position of the galaxy.  This makes Antlia similar to LGS~3, another
out-lying low surface brightness galaxy which contains HI, and as with
LGS~3, Antlia shows little evidence of recent star formation.

\acknowledgements{ }

\newpage

\def\x{\enspace}
\def\xx{\enspace\enspace}
\def\xxx{\enspace\enspace\enspace}
\def\xxxx{\enspace\enspace\enspace\enspace}
\def\xxxxx{\enspace\enspace\enspace\enspace\enspace}
\def\jref#1 #2 #3 #4 {{\par\noindent \hangindent=3em \hangafter=1
      \advance \rightskip by 5em #1, {\it#2}, {\bf#3}, {#4} \par}}
\def\ref#1{{\par\noindent \hangindent=3em \hangafter=1
      \advance \rightskip by 5em #1 \par}}
\def\endtable{\endgroup}
\def\tableheight{\vrule width 0pt height 8.5pt depth 3.5pt}
{\catcode`|=\active \catcode`&=\active
    \gdef\tabledelim{\catcode`|=\active \let|=\vbar
                     \catcode`&=\active \let&=\nobar} }
\def\table{\begingroup
    \def\twidth{\hsize}
    \def\tablewidth##1{\def\twidth{##1}}
    \def\defaultheight{\vrule width 0pt height 8.5pt depth 3.5pt}
    \def\heightdepth##1{\dimen0=##1
        \ifdim\dimen0>5pt
            \divide\dimen0 by 2 \advance\dimen0 by 2.5pt
            \dimen1=\dimen0 \advance\dimen1 by -5pt
            \vrule width 0pt height \the\dimen0  depth \the\dimen1
        \else  \divide\dimen0 by 2
            \vrule width 0pt height \the\dimen0  depth \the\dimen0 \fi}
    \def\spacing##1{\def\defaultheight{\heightdepth{##1}}}
    \def\nextheight##1{\noalign{\gdef\tableheight{\heightdepth{##1}}}}
    \def\end{\cr\noalign{\gdef\tableheight{\defaultheight}}}
    \def\zerowidth##1{\omit\hidewidth ##1 \hidewidth}
    \def\hline{\noalign{\hrule}}
    \def\skip##1{\noalign{\vskip##1}}
    \def\bskip##1{\noalign{\hbox to \twidth{\vrule height##1 depth 0pt \hfil
        \vrule height##1 depth 0pt}}}
    \def\header##1{\noalign{\hbox to \twidth{\hfil ##1 \unskip\hfil}}}
    \def\bheader##1{\noalign{\hbox to \twidth{\vrule\hfil ##1
        \unskip\hfil\vrule}}}
    \def\spanloop{\span\omit \advance\mscount by -1}
    \def\extend##1##2{\omit
        \mscount=##1 \multiply\mscount by 2 \advance\mscount by -1
        \loop\ifnum\mscount>1 \spanloop\repeat \ \hfil ##2 \unskip\hfil}
    \def\vbar{&\vrule&}
    \def\nobar{&&}
    \def\hdash##1{ \noalign{ \relax \gdef\tableheight{\heightdepth{0pt}}
        \toks0={} \count0=1 \count1=0 \putout##1\end
        \toks0=\expandafter{\the\toks0 &\end} \xdef\piggy{\the\toks0} }
        \piggy}
    \let\e=\expandafter
    \def\putspace{\ifnum\count0>1 \advance\count0 by -1
        \toks0=\e\e\e{\the\e\toks0\e&\e\multispan\e{\the\count0}\hfill}
        \fi \count0=0 }
     \def\putrule{\ifnum\count1>0 \advance\count1 by 1
        \toks0=\e\e\e{\the\e\toks0\e&\e\multispan\e{\the\count1}\leaders\hrule\
hfill}
        \fi \count1=0 }
    \def\putout##1{\ifx##1\end \putspace \putrule \let\next=\relax
        \else \let\next=\putout
            \ifx##1- \advance\count1 by 2 \putspace
            \else    \advance\count0 by 2 \putrule \fi \fi \next}   }
\def\tablespec#1{
    \def\vdimens{\noexpand\tableheight}
    \def\tabby{\tabskip=0pt plus100pt minus100pt}
    \def\r{&################\tabby&\hfil################\unskip}
    \def\c{&################\tabby&\hfil################\unskip\hfil}
    \def\l{&################\tabby&################\unskip\hfil}
    \edef\templ{\noexpand\vdimens ########\unskip  #1
         \unskip&########\tabskip=0pt&########\cr}
    \tabledelim
    \edef\body##1{ \vbox{
        \tabskip=0pt \offinterlineskip
        \halign to \twidth {\templ ##1}}} }

\centerline{\bf Table 1: The Galaxy Sample}
\vskip-1cm
$$
\table
\tablespec{\l\c\c\l\l}
\body{
\skip{0.06cm}
\hline
\skip{0.025cm}
\hline
\skip{.2cm}
& Object & Distance & M$_V$ & type & ref &\end
&  & ~~(kpc) &  &  & &\end
\skip{.1cm}
\hline
\skip{.05cm}
\hline
\skip{.45cm}
& Antlia & 1235&$-$10.8 & dIrr/dSph & Whiting {\it et al.} 1997 &\end
&        &     &        &           & Fouqu\'{e} {\it et al.} 1990 &\end
& Cetus & 800 & $-$10.1 & dSph & Whiting {\it et al.} 1999 &\end
\skip{.3cm}
& Tucana & 880 & $-$9.6 & dSph & Lavery \& Mighell 1992 &\end
\skip{.25cm}
\hline
\skip{0.025cm}
\hline
}
\endtable
$$
\vskip2cm
\newpage
\centerline{\bf Table 2: The Observations}
\vskip-1cm
$$
\table
\tablespec{\l\l\l\r\c\c\l}
\body{
\skip{0.06cm}
\hline
\skip{0.025cm}
\hline
\skip{.2cm}
& Date & Begin &Object & Exptime & Airmass & Seeing & Comments &\end
& & UT & & ~~(secs) & & ~(arcsec) &&\end
\skip{.1cm}
\hline
\skip{.05cm}
\hline
\skip{.45cm}
& 11May99 & 23:53 & Rup~106--1 & 150  & 1.2  &1.3       &Light Cirus&\end
& 12May99 & 00:12 & HD~80170--1      &   1  & 1.1  &0.85      &&\end
& & 00:29 & Antlia--1 & 5400 & 1.1&0.7$-$1.4 &Rapidly deteriorating seeing&\end
& & 02:11 & HD~92588      &   1  & 1.3  &1.9       &&\end 
\skip{.1cm}
\skip{.1cm}
& 6June99 & 00:04 & Antlia--2 & 5400&1.2&0.7$-$1.6 &Highly variable seeing&\end
& & 01:51 & Rup~106--2 &  150 & 1.2  &1.5       &&\end
& & 02:13 & HD~80170--2      &    1 & 2.4  &0.95      &&\end
& & 02:25 & HD~107328     &    1 & 1.4  &0.9       &&\end
& & 02:48 & Rup~106--3 &  150 & 1.3  &1.1       &&\end
\skip{.1cm}
\skip{.1cm}
& 26July99 & 07:16 & Pal~12 & 150 & 1.0 & 1.0 & Cloudy, full Moon &\end
& & 07:32 & HD~203638 & 1 & 1.1 & 1.0 & data too poor to use &\end
\skip{.1cm}
\skip{.1cm}
& 24Aug99 & 05:14 & HD~223647--1 &1&1.9  &1.25 &Full Moon and thick cirrus&\end
& & 05:40 & NGC~6752--1    &  250 & 1.7  &1.15      &&\end
& & 06:10 & Tucana--1   & 5400 & 1.4  &1.1$-$1.7 &&\end
& & 07:56 & HD~223647--2   &    6 & 1.9  &1.85      &&\end
& & 08:13 & Tucana--2   & 5400 & 1.7  &1.3$-$2.1 &&\end
& & 09:59 & HD~223647--3    &    6 & 2.0  &1.5       &&\end
\skip{.1cm}
\skip{.1cm}
& 26Aug99 & 03:49 & NGC~6752--2     &  250 & 1.7  &0.5       &Full Moon &\end
& & 04:09 & HD~223647--4    &    1 & 1.9  &0.5       &&\end
& & 04:11 & HD~223647--5    &    1 & 1.9  &0.5       &&\end
& & 04:36 & Cetus--1    & 5400 & 1.1  &0.5$-$0.8 &Seeing deteriorating&\end
& & 06:31 & HD~693--1      &    1 & 1.0  &0.7       &&\end
& & 06:34 & HD~693--2      &    1 & 1.0  &0.7       &&\end
& & 06:46 & Cetus--2    & 5400 & 1.07 &0.55$-$1.1&&\end
& & 08:32 & HD~8779--1     &    1 & 1.1  &0.9       &&\end
& & 08:35 & HD~8779--2     &    1 & 1.1  &0.9       &&\end
\skip{.25cm}
\hline
\skip{0.025cm}
\hline
}
\endtable
$$
\newpage

\renewcommand{\thefootnote}{\fnsymbol{footnote}}

\centerline{\bf Table 3: The Calibrators}
\vskip-1cm
$$
\table
\tablespec{\l\l\c\r\l}
\body{
\skip{0.06cm}
\hline
\skip{0.025cm}
\hline
\skip{.2cm}
& Object & Class &V & v$_{\odot}$ &\end
&  & && ~~km/s & Ref &\end
\skip{.1cm}
\hline
\skip{.05cm}
\hline
\skip{.45cm}
& Rup~106& Cluster & 10.9 & $-$44.0 & Da Costa {\it et al.} 1992&\end
& NGC~6752    & Cluster & 5.4 & $-$27.9 & Harris 1996&\end
& Pal~12      & Cluster & 11.99 & +27.8 & Harris 1996&\end
\skip{.2cm}
\skip{.2cm}
& HD~80170 & K5 III-IV& 5.33 & +0.0 &&\end
& HD~92588 & K1 IV& 6.26 & +42.8 &&\end
& HD~107328& K1 III & 4.96 & +35.7 &&\end
& HD~223647& G5 III & 5.11 & +13.8 &&\end
& HD~693\footnotemark[2]& F6 V& 
4.89 & +14.7 &&\end
& HD~8779  & K0 III & 6.41 & $-$5.0 &&\end
\skip{.25cm}
\hline
\skip{0.025cm}
\hline
}
\endtable
$$
\footnotetext[2]{Not used due to large template mismatch} 
\newpage

\centerline{\bf Table 4: Individual Results}
\vskip-1cm
$$
\table
\tablespec{\l\r\r\r\r\r}
\body{
\skip{0.06cm}
\hline
\skip{0.025cm}
\hline
\skip{.2cm}
& Object & v$_{obs}$ & Temp. Corr & Helio Corr. & Flex. Corr & v$_r$ &\end
&  & ~~(km/s) &  ~~(km/s)& ~~(km/s) &~~(km/s) &~~(km/s)&\end
\skip{.1cm}
\hline
\skip{.05cm}
\hline
\skip{.45cm}
& HD~107328    & +0.0& +63.3 & $-$27.6& +0.0 &+35.7 &\end
& HD~80170--1  & $-$30.1& +63.3 & $-$17.2 & $-$7.9 & +8.1&\end
& HD~80170--2  & +4.2  & +63.3 & $-$18.2 & $-$35.5 & +13.8 &\end    
& HD~92588     & +1.3  & +63.3 & $-$26.9 & $-$3.9 &  +33.8&\end
& HD~223647--1 & +18.9 & +63.3& $-$8.6& $-$55.3  & +18.3 &\end
& HD~223647--2 & +19.9 & +63.3& $-$8.6& $-$61.0  & +13.6 &\end
& HD~223647--3 & +11.9 & +63.3& $-$8.6& $-$59.2  & +7.4 &\end
& HD~223647--4 & +11.9 & +63.3& $-$8.6& $-$35.5 & +31.1&\end
& HD~223647--5 & $-$1.2& +63.3& $-$8.6& $-$23.7 & +29.8&\end
& HD~8779--1   & $-$58.3&+63.3& +21.2 & $-$11.8 & +14.4& \end
& HD~8779--2   & $-$61.2&+63.3& +21.2 & $-$7.9 & +15.4& \end
\skip{.1cm}
\skip{.1cm}
& NGC~6752--1 & -45.1 &    +63.3  &   $-$18.5  &  -19.0  &   $-$19.3&\end
& NGC~6752--2 & $-$53.3& +63.3& $-$18.5& $-$11.9 & $-$20.4&\end
\skip{.1cm}
\skip{.1cm}
& Rup~106--1-a &+250.5 &   +63.3 & $-$6.8 & $-$19.7&+287.3&\end
& Rup~106--1-b & +5.7 &   +63.3 & $-$6.8 & $-$19.7& +42.5&\end
& Rup~106--1-c & $-$11.4&   +63.3 & $-$6.8 & $-$19.7&+25.4&\end
& Rup~106--1-d &  $-$50.9&   +63.3 & $-$6.8 & $-$19.7& $-$14.1&\end
& Rup~106--1-e &  $-$87.1&   +63.3 & $-$6.8 & $-$19.7& $-$50.4&\end
& Rup~106--1-unresolved&  $-$50: &   +63.3 & $-$6.8 & $-$19.7& $-$14:&\end
\skip{.1cm}
\skip{.1cm}
& Rup~106--2-a &+188.9& +63.3 & $-$14.2&+7.9 &+245.9&\end
& Rup~106--2-b &$-$34.4& +63.3 & $-$14.2&+7.9 & +22.6&\end
& Rup~106--2-c &$-$28.0& +63.3 & $-$14.2&+7.9 &  +29.0&\end
& Rup~106--2-d &$-$53.1& +63.3 & $-$14.2&+7.9 & +3.9&\end
& Rup~106--2-e &$-$57.7& +63.3 & $-$14.2&+7.9 & $-$0.7&\end
\skip{.1cm}
\skip{.1cm}
& Rup~106--3-w & +27.3& +63.3& $-$14.2& $-$11.8 &+64.6&\end
& Rup~106--3-x & $-$10.9& +63.3& $-$14.2& $-$11.8 &+26.4&\end
& Rup~106--3-y & $-$18.9& +63.3& $-$14.2& $-$11.8 &+18.4&\end
& Rup~106--3-z & $-$84.1& +63.3& $-$14.2& $-$11.8 & $-$46.8&\end
\skip{.1cm}
\skip{.1cm}
&Antlia--1-a   & $-$41.8&    +63.3& $-$21.6&+4.7 & $-$4.6&\end
&Antlia--1-b   &     +353.2&    +63.3& $-$21.6&+4.7 &+399.6&\end
&Antlia--1-c  &     +308.1&    +63.3& $-$21.6&+4.7 &+354.5&\end
&Antlia--1-d   &    +300:  &    +63.3& $-$21.6&+4.7 &+346:&\end
&Antlia--1-e   &     $-$142: &    +63.3& $-$21.6&+4.7 & $-$96:&\end
&Antlia--1-unresolved &    +380:  &    +63.3& $-$21.6&+4.7 & +426:&\end
\skip{.1cm}
\skip{.1cm}
&Antlia--2-b   & +329.1 & +63.3& $-$23.6&+0.5&+371.1&\end
&Antlia--2-c   & +303.8 & +63.3& $-$23.6&+0.5&+344.0&\end
&Antlia--2-d   & +318.1 & +63.3& $-$23.6&+0.5&+358.3&\end
&Antlia--2-e   &  $-$183.6& +63.3& $-$23.6&+0.5& $-$143.4&\end
&Antlia--2-unresolved & +366.9 & +63.3& $-$23.6&+0.5& +407.1&\end
\skip{.1cm}
\hline
\skip{.05cm}
\hline
}
\endtable
$$
\newpage

\centerline{\bf Table 5: Summary of Results}
\vskip-1cm
$$
\table
\tablespec{\l\c\r\c}
\body{
\skip{0.06cm}
\hline
\skip{0.025cm}
\hline
\skip{.2cm}
& Object    & $<$v$_r$ obs$>$ & v$_r$ {known} & Comments &\end
&           & ~~(km/s)    &  ~~(km/s)   &          &\end
\skip{.1cm}
\hline
\skip{.05cm}
\hline
\skip{.45cm}
& HD~107328 & $+35.7\pm0.0$\footnotemark[1]& $+35.7\pm0.3$  & RV template&\end
& HD~80170  & $+10.9\pm7.8$ & $+0.0\pm0.2$   & RV standard&\end
& HD~92588  & $+33.8\pm6.9$ & $+42.8\pm0.1$  & RV standard&\end
& HD~223647 & $+20.0\pm7.4$ & $+13.8\pm0.4$  & RV standard&\end
& HD~8779   & $+14.9\pm8.3$ & $-5.0\pm0.6$  & RV standard&\end
\skip{.2cm}
\skip{.2cm}
& Rup~106   &   :\footnotemark[2]& $-44.0\pm3.0$  & Cluster Standard&\end
& NGC~6752  & $-19.8\pm6.7$ & $-27.9\pm0.8$  & Cluster Standard &\end
\skip{.2cm}
\skip{.2cm}
&Antlia--1-bcd &$363.1\pm16.4$& $361\pm2$& HI velocity from Fouqu\'{e} {\it et 
al.}&\end
&Antlia--2-bcd &$335.3\pm17.1$& & &\end
&Antlia--1,2-bcd &$349.5\pm11.7$& &  HD~107328 as template&\end
&Antlia--1,2-bcd &$351.4\pm12.9$& &  NGC6752 as template&\end
\skip{.25cm}
\hline
}
\endtable
$$
\footnotetext[1]{This is the rv primary standard, and so by definition
has a relative rv of 0.0 with no error.} 
\footnotetext[2]{Not useable, due to poor s/n, and a very crowded
field of view}
\newpage

\begin{figure}
\centerline{\hbox{\psfig{figure=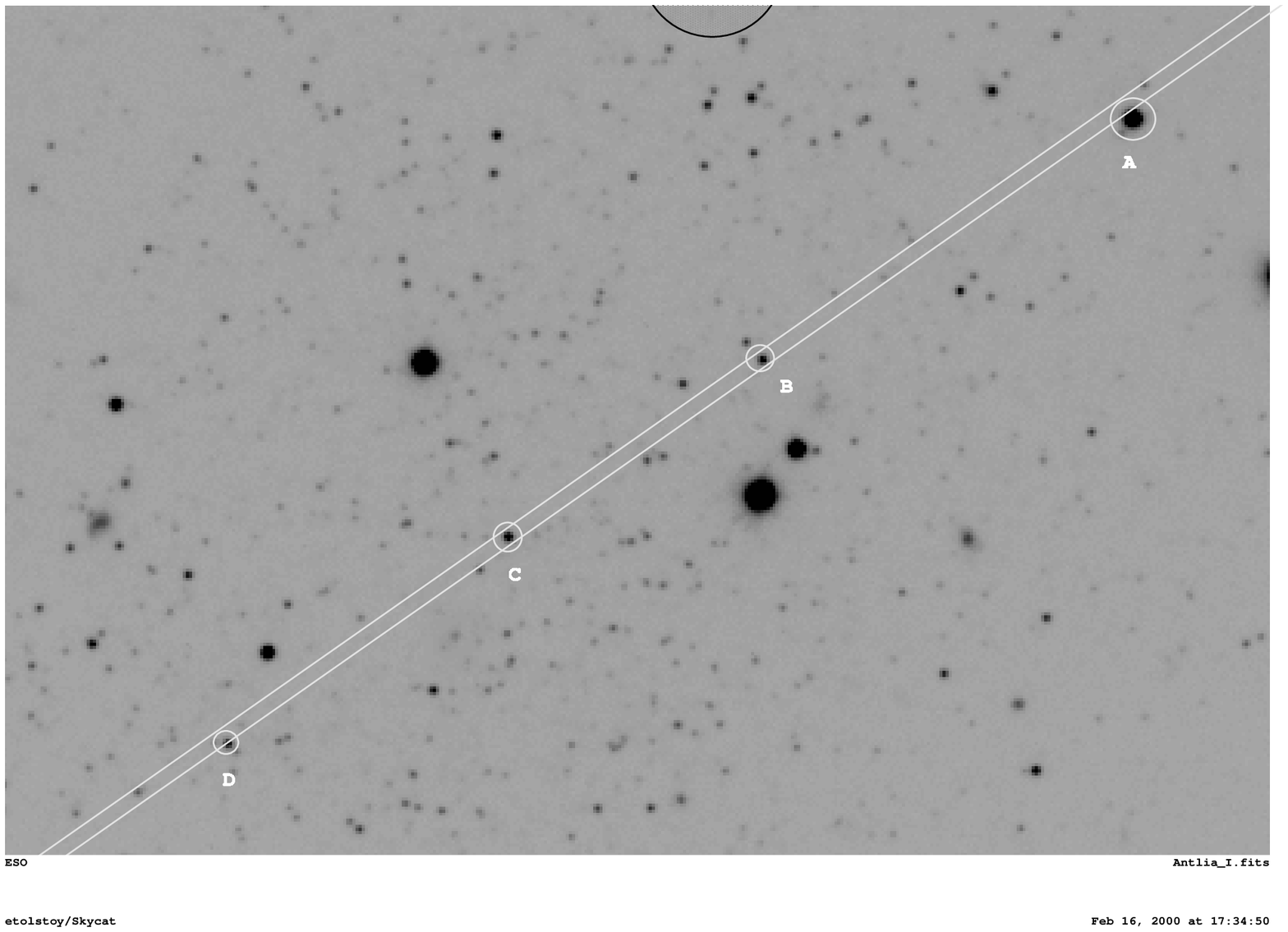,width=18cm}}}
\caption{The central 4 arcmin of the Antlia galaxy
with the adopted slit position marked. North is up and East is left.
This plot comes
from the FIMS, obervation preparation tool output. The image
is a combined 5400sec of I filter observations made with FORS1
for FORS1 Science Verification in January 1999.
Marked on the image 
are 4 of the 5 stars indentified in our slit and for which
individual spectra could be usefully extracted. }
\label{first}
\end{figure}
\newpage
\begin{figure}
\centerline{\hbox{\psfig{figure=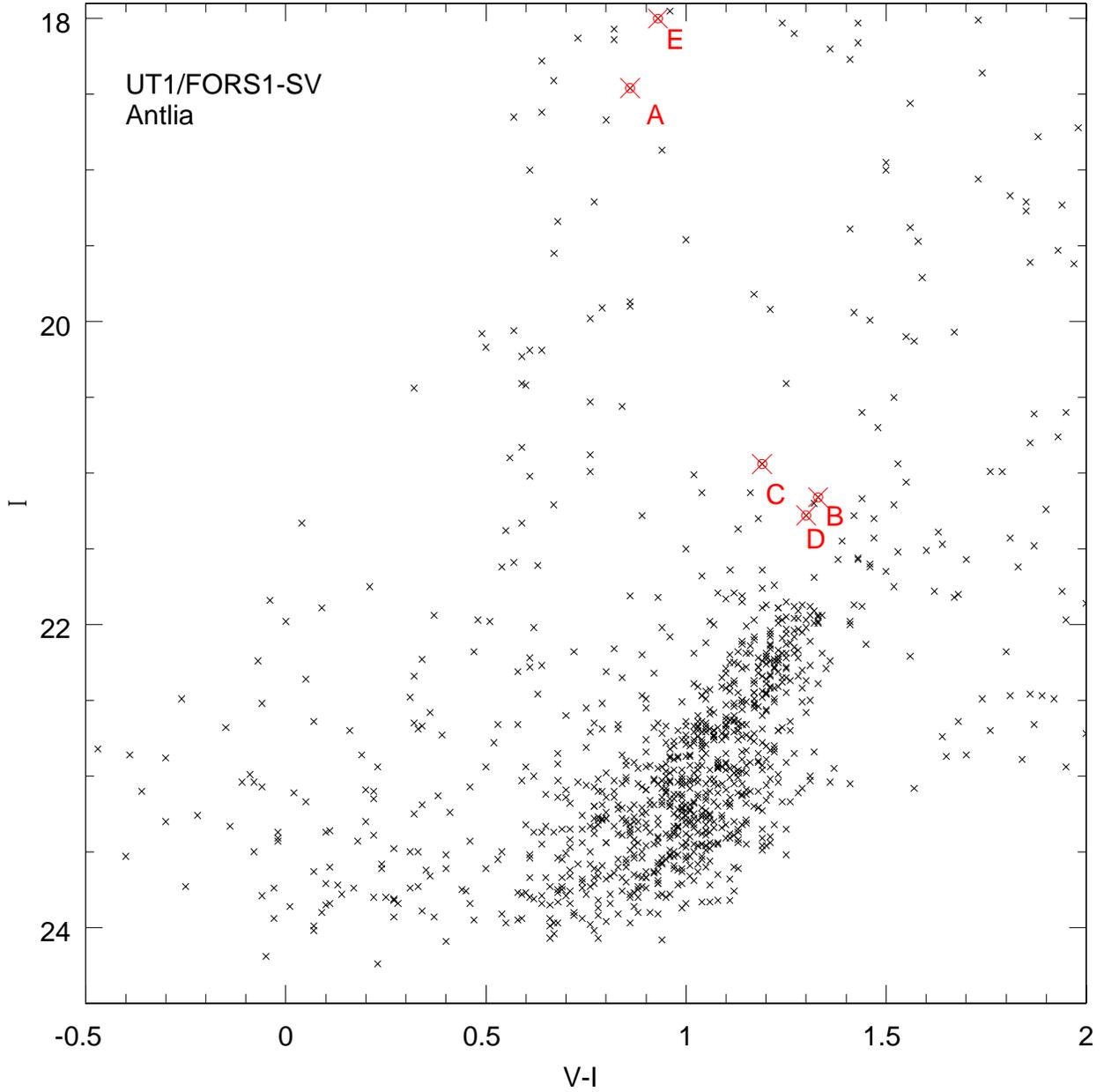,width=18cm}}}
\caption{An Antlia (I, V-I) Colour-Magnitude
Diagram made from FORS1 Science Verification data taken in
January 1999. Plotted in large cross symbols are stars
in the slit for which we could extract individual spectra.
The labels correspond to the letters in Figure~1 and in
Table~4.  Objects B, C, D lie near the tip of the red giant branch
and have similar velocities (see Table~4).}
\label{second}
\end{figure}
\newpage
\begin{figure}
\centerline{\hbox{\psfig{figure=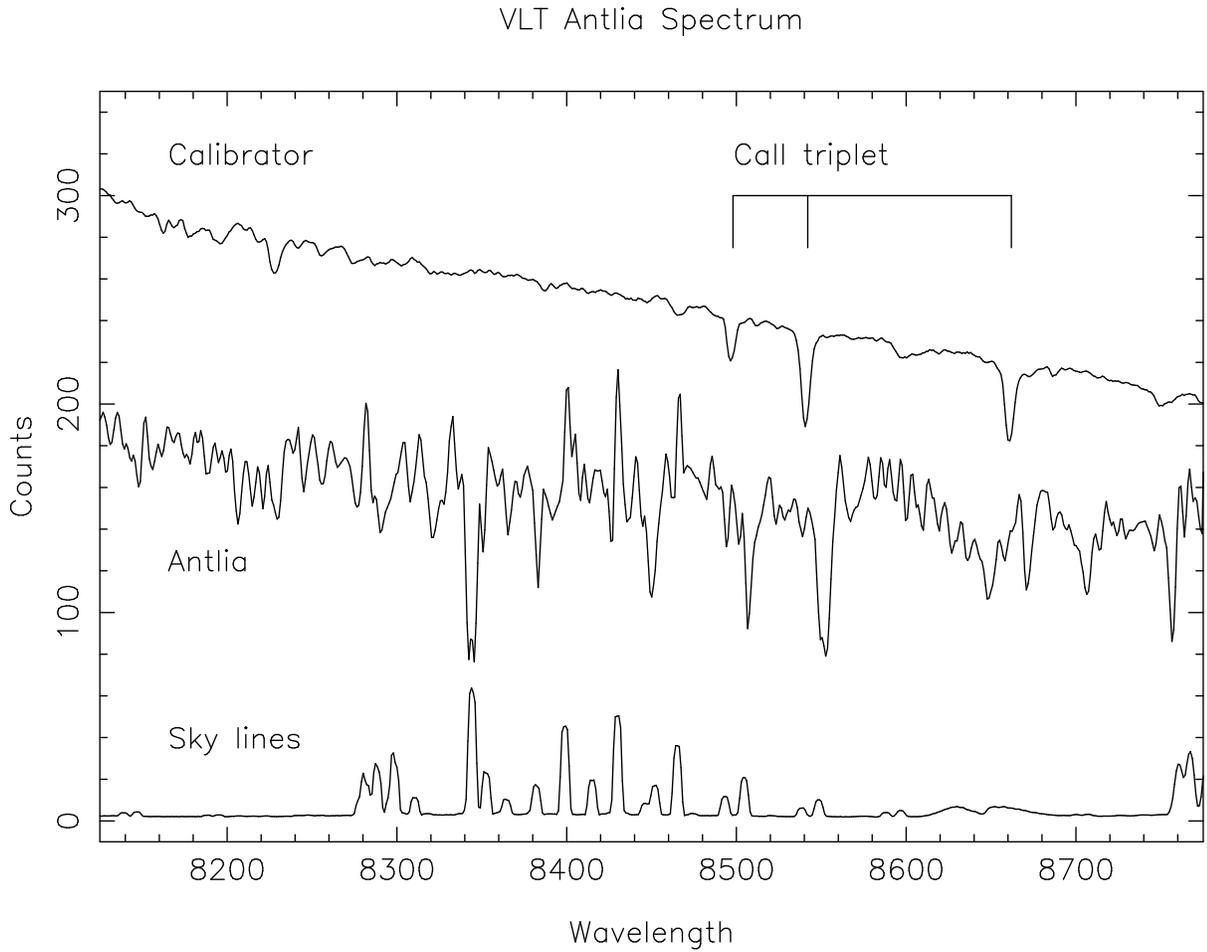,width=18cm,angle=-90}}}
\caption{Here we plot the combined Antlia spectrum of objects b, c and d
from both observation epochs.  Also plotted for comparison above
Antlia is the spectrum of the globular cluster 
spectrum of NGC~6752 (Calibrator).  The
Ca~II triplet lines for Antlia are clearly redshifted by $\approx$10\AA \ 
with respect to the calibrator cluster.  
Also included at the bottom of the plot
is a scaled down (by a factor of 75) sky spectrum illustrating the problems 
of residual sky line contamination.  }
\label{third}
\end{figure}
\newpage
\begin{figure}
\centerline{\hbox{\psfig{figure=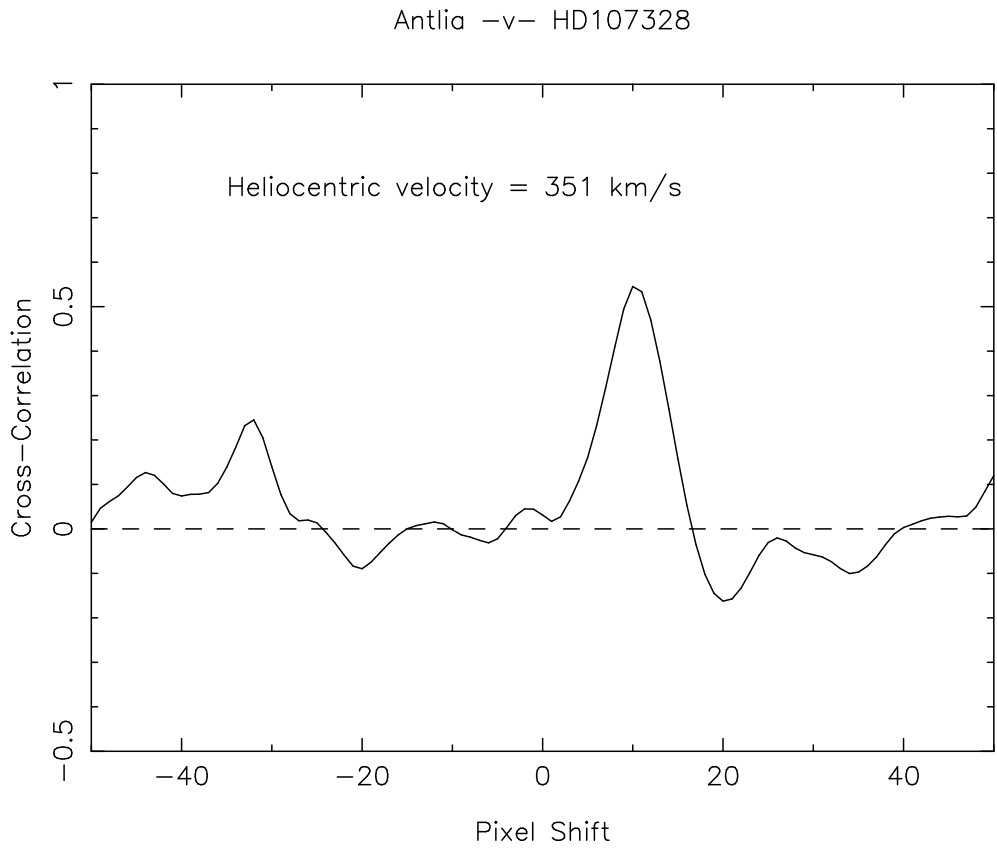,width=20cm,angle=-90}}}
\caption{The cross-correlation of the Antlia spectrum,
shown in Figure~3, against the HD~107328 template spectrum.  No template,
flexure or helio-centric corrections have been made for this plot. }
\label{fourth}
\end{figure}

\end{document}